\documentclass[sigconf]{acmart}
\usepackage{multirow}

\AtBeginDocument{%
  }

\setcopyright{acmlicensed}
\copyrightyear{2026}
\acmYear{2026}
\acmDOI{XXXXXXX.XXXXXXX}
\acmConference[Conference acronym 'XX]{Make sure to enter the correct
  conference title from your rights confirmation email}{June 03--05,
  2018}{Woodstock, NY}
\acmISBN{978-1-4503-XXXX-X/2018/06}

\begin{document}

\title{SEEK: Skill-Routed Evaluation with Evolvable Knowledge for Industrial Search}

\author{Zhongxin Huang}
\authornote{Equal contribution.}
\email{huangzhongxin26@stu.pku.edu.cn}
\affiliation{%
  \institution{Peking University}
  \city{Beijing}
  \country{China}
}

\author{Songyang Li}
\authornotemark[1]                    
\email{lisongyang03@kuaishou.com}
\affiliation{%
  \institution{Kuaishou Technology}
  \city{Beijing}
  \country{China}
}

\author{Renzhe Zhou}
\authornotemark[1]
\email{zhourenzhe03@kuaishou.com}
\affiliation{%
  \institution{Kuaishou Technology}
  \city{Hangzhou}
  \country{China}
}

\author{Feiran Zhu}
\authornotemark[1]
\email{a499616042@163.com}
\affiliation{%
  \institution{Unaffiliated}
  \city{Hangzhou}
  \country{China}
}

\author{Chenglei Dai}
\email{daichenglei@kuaishou.com}
\affiliation{%
  \institution{Kuaishou Technology}
  \city{Hangzhou}
  \country{China}
}

\author{Zhen Xiao}
\authornotemark[2]
\email{xiaozhen@pku.edu.cn}
\affiliation{%
  \institution{Peking University}
  \city{Beijing}
  \country{China}
}

\author{Xuanping Li}
\email{lixuanping@kuaishou.com}
\affiliation{%
  \institution{Kuaishou Technology}
  \city{Beijing}
  \country{China}
}

\author{Jingwei Zhuo}
\authornote{Corresponding author.}      
\email{zhuojw10@gmail.com}
\affiliation{%
  \institution{Unaffiliated}
  \city{Beijing}
  \country{China}
}

\renewcommand{\shortauthors}{Huang et al.}

\begin{abstract}

Search quality evaluation provides essential supervision and diagnostic
signals for the development and iteration of industrial search systems.
Although large language models (LLMs) offer a scalable alternative to
manual assessment, reliable automatic evaluation remains challenging:
users experience search results at the page level, while the applicable
evaluation criteria are multi-dimensional and continuously evolving.
Packing all evaluation criteria into a unified prompt introduces
irrelevant context and potential criterion interference, whereas
internalizing them through post-training tightly couples
rule updates with costly model retraining cycles.

To address these issues, we propose Skill-routed Evaluation with Evolvable Knowledge (SEEK).
Specifically, SEEK externalizes specific search evaluation criteria into a skill bank,
dynamically routes relevant skills for each
query--result list pair, and
employs a task-adapted listwise evaluator to produce
page-level judgments and failure mode attribution.
A two-stage training pipeline teaches the evaluator to align evaluation criteria with human preferences, while a replay-gated skill bank allows recurring evaluation knowledge gaps to be incorporated without model retraining.
Experiments on industrial short-video search show that SEEK improves listwise quality evaluation accuracy and achieves significant progress in attribution diagnosis.
SEEK has been deployed at Kuaishou, a short-video platform with over 400 million daily active users, significantly improving the scale and quality of online search evaluation.
\end{abstract}




\maketitle

\section{Introduction}

Search quality evaluation is a fundamental component of modern search systems.
It provides the labels and diagnostic signals required for model
development, offline comparison and online experimentation~\cite{dewan2025llm,li2024llms}.
Traditionally, such evaluations rely on trained human assessors.
Although human judgments remain the most reliable source of
task-specific supervision, their high cost and long turnaround time make it
difficult to cover rapidly changing traffic, long-tail queries, and
frequent system iterations at industrial scale~\cite{tang2025lref,lu2025vlm,lu2025lore}.

Large language models (LLMs) offer a promising alternative for scaling search evaluation.
Recent studies have shown that carefully calibrated LLMs can generate useful relevance judgments and achieve agreement with human assessors on both academic and industrial search tasks
~\cite{thomas2024large,upadhyay2024umbrela,wang2025llm,zeng2026optimizing}.
These results have motivated a growing body of work on using LLMs as automatic judges for retrieval and search systems~\cite{balog2025rankers,sharifymoghaddam2025rankllm,zhu2025large,zhai2024large}.
However, directly prompting a general-purpose LLM does not fully satisfy the requirements of industrial search evaluation.

One challenge is that users experience a search result page as a whole rather than as isolated retrieved items.
Page quality depends jointly on relevance, ranking, content quality, redundancy, intent coverage, and source authority
~\cite{lahiri2026design,levine2025relevance,guan2025faithfulness,zhang2026efficient,dewan2026llm}. Several of these properties are inherently page-level. Recent work has begun to move beyond independent query--item grading through batched relevance assessment, usefulness-oriented
rubrics, and behavior-grounded evaluation
~\cite{korikov2025batched,dewan2026llm,vardasbi2026aligning}. Nevertheless, most existing automatic search evaluation methods still focus primarily on item-level relevance or ranking ~\cite{liu2026bridging,wang2025clue}, rather than jointly predicting page-level quality and causes of quality degradation.

Another challenge is how to represent rich and continuously evolving evaluation knowledge.
Packing all rules, boundary cases, and evidence requirements into a
single prompt exposes every sample to largely irrelevant criteria,
creating contextual redundancy and potential criterion interference
~\cite{baysan2025llm,arabzadeh2025human,dun2025sweeping,chen2026skill}.
Task-specific supervised fine-tuning (SFT) and reinforcement learning (RL) have
been increasingly adopted to specialize LLMs for industrial Web
applications, including search relevance modeling and sequential
decision making
~\cite{tang2025lref,lu2025lore,yang2025taosr,song2026dara}.
However, they also internalize evaluation knowledge into model parameters.
As platform content, query intents, and evaluation policies evolve,
updating such knowledge may require another cycle of data construction,
training, validation, and deployment
~\cite{zhao2026survey,lu2025vlm}.
This creates a practical tension between keeping evaluation knowledge
rapidly updatable and keeping the underlying evaluation model stable.

To address these challenges, we propose Skill-routed Evaluation with Evolvable Knowledge (SEEK), a
skill-grounded framework that separates explicit evaluation knowledge
from model selection and reasoning capabilities.
SEEK externalizes domain-specific criteria into a modular skill bank
and uses a lightweight \emph{skill router} to select a compact subset of
task-relevant skills for each query--list pair.
The corresponding operational guidance is then provided, together with
the original result page, to a task-adapted \emph{listwise evaluator},
which produces a page-level judgment and structured attribution.
A human-aligned training pipeline transforms human judgments into
routing supervision and evidence-grounded evaluation trajectories,
enabling the router and evaluator to learn how to select and faithfully
execute the externalized knowledge.
Finally, the \emph{self-evolving skill bank} attributes reviewed
production failures to routing, knowledge, or execution errors and
incorporates recurrent knowledge gaps through localized,
replay-validated skill revisions.
This allows newly emerging evaluation knowledge to be updated.

Our main contributions are summarized as follows:

\begin{itemize}

    \item We propose SEEK, a skill-grounded framework for
    listwise search evaluation that decouples task-specific
    evaluation knowledge from model capabilities and dynamically routes
    relevant skills for page-level judgment.

    \item We introduce a human-aligned training strategy that
    transforms human search judgments into routing supervision and
    evaluation trajectories, enabling effective specialization
    of a compact listwise evaluator.

    \item We develop a self-evolving skill bank that updates
    reusable evaluation knowledge through error attribution and
    replay-gated revision, enabling lightweight adaptation without
    immediate model retraining.

    \item Extensive offline and online experiments demonstrate
    the effectiveness and practical value of SEEK. The system has
    been deployed in Kuaishou's search evaluation pipeline.

\end{itemize}

\begin{figure*}[t]
    \centering
    \includegraphics[width=\textwidth]{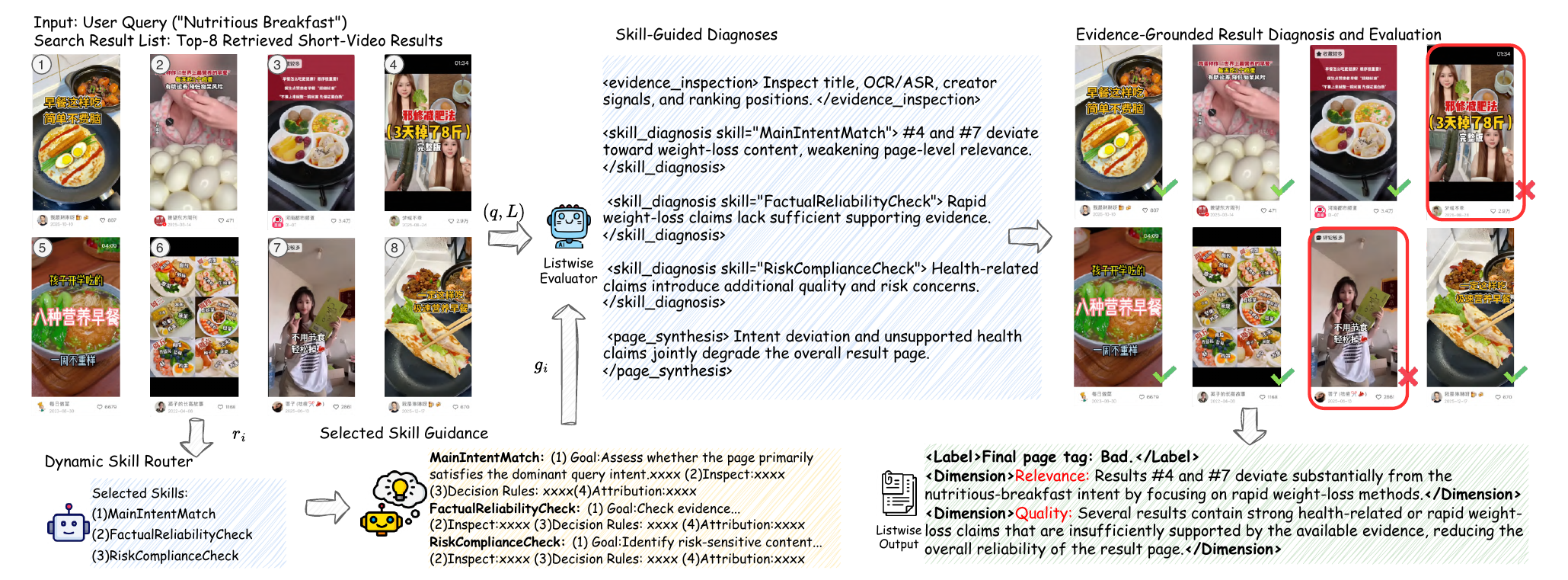}
    \caption{
    Illustrative example of skill-routed listwise evaluation in SEEK.
    The router selects task-relevant skills, and the listwise evaluator
    applies their guidance over the complete result page to produce a
    page-level judgment and attribution.
    }
    \label{fig:seek_example}
\end{figure*}

\section{Related Work}
\label{sec:related_work}

\subsection{LLM-based Search Evaluation}

Large language models have been increasingly explored as automatic
evaluators for search systems, covering relevance judgment,
usefulness assessment, and broader search-quality evaluation
~\cite{yang2025taosr,zeng2026optimizing,wang2025clue,lu2025vlm}.
Within this line of research, LLM-as-a-Judge has emerged as a
promising paradigm for replacing or assisting costly human assessment
in relevance evaluation
~\cite{lu2025vlm,rahmani2024llmjudge,balog2025rankers,farzi2025criteria}.
More recent studies further extend LLM-based evaluation toward
industrial objectives, including usefulness-oriented supervision,
fine-grained intent satisfaction, and web-scale relevance assessment,
substantially reducing annotation cost while improving evaluation
coverage and scalability
~\cite{dewan2025llm,choi2025bloomintent,wang2025llm}.

Despite this progress, most existing approaches either rely on
predefined evaluation criteria or internalize task-specific knowledge
through model post-training.
As search policies, content ecosystems, and evaluation standards
continuously evolve, such designs provide limited flexibility for
rapidly incorporating new evaluation knowledge without modifying the underlying evaluator.

\subsection{Listwise Evaluation}

Traditional IR evaluation typically aggregates item-level relevance
judgments into ranking metrics such as Precision, MAP, and nDCG.
Although these metrics are effective for measuring relevance and
ranking quality, they do not explicitly model interactions among
retrieved results, such as redundancy, intent coverage, and source
composition.
Recent work in RAG evaluation, listwise reranking, and whole-page
assessment has therefore increasingly moved beyond independent
item-level judgments toward jointly modeling the retrieved list as a
whole
~\cite{trappolini2026redefining,reddy2024first,
korikov2025batched,lahiri2026design,vardasbi2026aligning}.

However, industrial search evaluation requires not only an overall
page-level judgment, but also fine-grained diagnosis of the underlying
quality issues.
This is particularly important in short-video search, where the final
experience depends jointly on heterogeneous content signals, ranking
positions, and interactions across retrieved results.

\subsection{Externalized and Evolving Skills}

Recent research has increasingly explored decoupling task-specific
knowledge from model parameters and representing it through modular,
reusable structures.
Human judgment criteria, policies, precedents, and task instructions
can be externalized as explicit knowledge interfaces or reusable
skills, while agent systems further acquire and accumulate such skills
from execution trajectories and interaction experience
~\cite{le2026sage,chen2026skill,wang2026skillx,
ma2026skillgen,wang2026skilllibrary}.

More recent studies extend static skill repositories toward continual
refinement and self-evolution, including revising textual skills from
execution feedback, optimizing external skills while keeping model
parameters fixed, and updating skill libraries through interaction
experience
~\cite{liu2026skillrevise,yang2026skillopt,huang2026skill}.
Despite this progress, controlled evolution of task-specific
evaluation knowledge under a relatively stable evaluator
remains underexplored.
This challenge is particularly important in industrial listwise search
evaluation, where evaluation criteria evolve over time and knowledge
updates must remain localized, reliable, and backward-compatible.

\section{Method}
\label{sec:methods}
 
\paragraph{Problem statement.}
Given a user query \(q\) and a ranked result list
\(L=[d_1,\ldots,d_K]\), we study structured listwise search quality evaluation.
Each result \(d_i\) contains the content and metadata available for
assessment, including textual signals, creator information, interaction
statistics, and ranking position.
The task is to jointly evaluate the complete result page:
\begin{equation}
f_{\mathrm{assess}}(q,L)=(l,\{a_j\}_{j=1}^{m}),
\label{eq:task_definition}
\end{equation}
where \(l\in\{\textit{good},\textit{fair},\textit{bad}\}\) denotes the
page-level quality label, and
\(\{a_j\}_{j=1}^{m}\) is a set of structured attribution.
Each \(a_j\) identifies a quality dimension and provides an
evidence-grounded diagnosis of the corresponding quality issue.
The quality dimensions are drawn from
\(\mathcal{A}=\{\textit{Relevance},\textit{Quality},
\textit{Diversity},\textit{Authority},
\textit{Heterogeneous}\}\).

This prediction is inherently listwise, as it depends not only on the
quality of individual results, but also on their ranking positions and
cross-result interactions, such as redundancy, intent coverage, and
source composition.
Figure~\ref{fig:seek_example} illustrates this setting with a
short-video search example.
For the query ``Nutritious Breakfast'', most retrieved results satisfy
the dominant intent, while several results shift toward rapid
weight-loss content and contain insufficiently supported health-related
claims.
A reliable evaluator must therefore go beyond assigning an overall
page-level label and further identify the specific dimensions
responsible for the degradation, such as relevance and content quality.


\subsection{SEEK Framework}
\label{sec:framework}

\begin{figure*}[t]
    \centering
    \includegraphics[width=\textwidth]{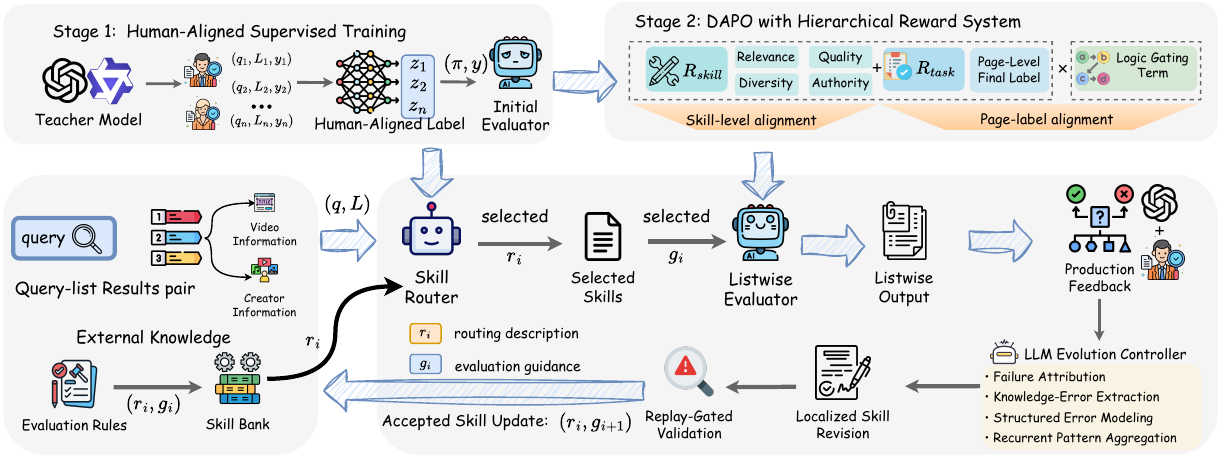}
    \vspace{-10pt}
    \caption{
    Overview of SEEK.
    The external skill bank stores explicit evaluation knowledge, while
    the skill router and listwise evaluator learn how to select and
    apply relevant knowledge for each query--list pair.
    Human-aligned supervision trains the two model components, whereas
    reviewed production feedback drives replay-gated evolution of the
    skill bank.
    }
    \Description{
    Overview of dynamic skill routing, skill-grounded listwise
    evaluation, and replay-gated evolution of external evaluation
    knowledge in SEEK.
    }
    \label{fig:framework}
\end{figure*}

As illustrated in Figure~\ref{fig:framework}, SEEK follows a capability--knowledge separation principle and consists of three
key components: the skill bank, skill router, and listwise evaluator.
The external skill bank stores explicit and evolvable evaluation
knowledge, while the skill router and listwise evaluator provide
relatively stable capabilities for selecting and applying such
knowledge, respectively.
Given a query--list pair \(x=(q,L)\), the router identifies a compact
set of applicable skills from the skill bank, and the corresponding
operational guidance is provided to the evaluator together with the
original query and result list.
The evaluator then performs skill-grounded listwise reasoning to
produce a page-level judgment and structured attribution.

\paragraph{Skill Bank.}
We maintain a skill bank
\(\mathcal{S}^{(t)}=\{S_1^{(t)},\ldots,S_N^{(t)}\}\), where \(t\)
denotes the current evolution round.
In our implementation, the bank contains 12 diagnostic skills spanning
five evaluation dimensions:
\textit{Relevance}, \textit{Quality}, \textit{Diversity},
\textit{Authority}, and \textit{Heterogeneous}.
Each skill is represented by two complementary textual components:
\begin{equation}
S_i^{(t)}
=
\left(
r_i^{(t)},\,g_i^{(t)}
\right),
\label{eq:skill}
\end{equation}
where \(r_i^{(t)}\) is a compact routing description specifying
\emph{when} the skill should be activated, and \(g_i^{(t)}\) is
structured operational guidance specifying \emph{how} the corresponding
criterion should be applied.
Concretely, \(g_i^{(t)}\) describes the evaluation objective, evidence
requirements, decision rules, boundary conditions, and attribution
requirements.
This separation allows routing and execution to rely on different
representations of the same evaluation criterion.

\paragraph{Skill Router.}
For each skill \(S_i^{(t)}\), the router estimates its applicability to
the current query--list pair using the corresponding routing
description:
\begin{equation}
p_i
=
R_{\phi}
\left(
q,L,r_i^{(t)}
\right),
\label{eq:router_score}
\end{equation}
where \(R_{\phi}\) denotes the router parameterized by \(\phi\), and
\(p_i\in[0,1]\) is the activation score of the \(i\)-th skill.
The active skill set is selected by thresholding these scores:
\begin{equation}
\mathcal{S}_{x}^{(t)}
=
\left\{
S_i^{(t)}\in\mathcal{S}^{(t)}
\mid
p_i>\tau
\right\},
\label{eq:skill_selection}
\end{equation}
where \(\tau\) is the activation threshold.
If no skill exceeds \(\tau\), the highest-scoring skill is retained as
a fallback.
In this way, the router exposes only task-relevant evaluation knowledge
to the downstream evaluator rather than injecting the entire skill bank
for every instance.

\paragraph{Listwise Evaluator.}
The evaluator receives both the original query--list pair and the
operational guidance associated with the selected skills:
\begin{equation}
(\widehat{\pi},\widehat{y})
=
E_{\theta}
\left(
q,L,
\left\{
g_i^{(t)}
\mid
S_i^{(t)}\in\mathcal{S}_{x}^{(t)}
\right\}
\right),
\label{eq:evaluator}
\end{equation}
where \(E_{\theta}\) denotes the listwise evaluator parameterized by
\(\theta\).
Rather than assessing retrieved items independently, the evaluator
jointly reasons over the complete ranked list under the selected
evaluation criteria.
It produces a structured evaluation trajectory
\(\widehat{\pi}\), which grounds the judgment in relevant evidence and
skill-level diagnoses, together with the final prediction
\(\widehat{y}=(\widehat{l},\{\widehat{a}_j\}_{j=1}^{\widehat{m}})\).
Here, \(\widehat{l}\) is the page-level quality label and
\(\{\widehat{a}_j\}_{j=1}^{\widehat{m}}\) denotes the corresponding
structured attribution.

\subsection{Skill-Aligned Training}
\label{sec:training}

The skill router and listwise evaluator learn two complementary
capabilities: the router determines which evaluation knowledge is
relevant, while the evaluator learns how to apply the selected
knowledge to the result page.
We train the two components separately using supervision derived from
the same human-annotated query--list pairs.
Human judgments provide the final evaluation targets, while a strong
teacher model transforms them into skill-routing labels and
evidence-grounded evaluation trajectories.

\subsubsection{Teacher-Guided Data Construction}
\label{sec:teacher_data}

Given a query--list pair \((q,L)\), the current skill bank
\(\mathcal{S}^{(t)}\), and the human reference
\(y^{*}=(l^{*},\mathcal{A}^{*})\), the teacher first identifies a
compact set of skills sufficient to support the annotated judgment.
It produces a multi-label routing target
\(z^{*}\in\{0,1\}^{N}\), which defines the reference skill set
\(\mathcal{S}_{x}^{*}
=\{S_i^{(t)} \mid z_i^{*}=1\}\).
The corresponding operational guidance is denoted as
\(\mathcal{G}_{x}^{*}
=\{g_i^{(t)} \mid S_i^{(t)}\in\mathcal{S}_{x}^{*}\}\).

Conditioned on the selected guidance and the human reference, we employ
a strong general-purpose LLM as the teacher model to generate a
structured and evidence-grounded evaluation trajectory:
\begin{equation}
\pi^{*}
=
T_{\mathrm{LLM}}
\left(
q,L,\mathcal{G}_{x}^{*},y^{*}
\right),
\label{eq:teacher_trajectory}
\end{equation}
where \(T_{\mathrm{LLM}}\) denotes the teacher model.
The teacher provides high-quality reasoning supervision by connecting
retrieved evidence to skill-level diagnoses and further synthesizing
these diagnoses into the human-annotated page-level judgment and attribution.

Each distilled instance is represented as
\((q,L,z^{*},\pi^{*},y^{*})\).
Before training, we filter samples with invalid skill references,
malformed trajectory structures, unsupported evidence, or explicit
inconsistencies with the human annotation.
The verified routing target \(z^{*}\) supervises the router, while
\((\pi^{*},y^{*})\) provides skill-conditioned supervision for the
evaluator.

\subsubsection{Router Training}

Because multiple evaluation criteria may be relevant to the same
query--list pair, we formulate skill routing as a multi-label
classification problem.
For \(N\) skills, the router is optimized using binary cross-entropy:
\begin{equation}
\mathcal{L}_{\mathrm{router}}
=
-\frac{1}{N}
\sum_{i=1}^{N}
\left[
z_i^{*}\log p_i
+
(1-z_i^{*})\log(1-p_i)
\right],
\label{eq:router_loss}
\end{equation}
where \(p_i\) is the predicted activation score for skill \(S_i\).
This objective directly aligns the router with the teacher-derived
skill-selection targets. We optimize the router independently from downstream reinforcement learning.


\subsubsection{Evaluator Training}

We train the listwise evaluator in two stages.
First, supervised fine-tuning teaches the model to reproduce verified
skill-conditioned evaluation trajectories and human judgments:
\begin{equation}
\mathcal{L}_{\mathrm{SFT}}
=
-
\log
P_{\theta}
\left(
\pi^{*},y^{*}
\mid
q,L,\mathcal{G}_{x}^{*}
\right).
\label{eq:sft}
\end{equation}
This stage establishes the basic capability to identify supporting
evidence, apply the selected evaluation criteria, generate
skill-grounded diagnoses, and synthesize them into a coherent
page-level decision.

We then apply DAPO~\cite{yu2025dapo} to further align the evaluator with human search
judgments.
For each query--list pair, we sample a group of candidate outputs
\(\mathcal{O}=\{o_1,\ldots,o_M\}\) and score each candidate using a
hierarchical reward:
\begin{equation}
R(o)
=
\left(
R_{\mathrm{task}}(o)
+
\lambda_s R_{\mathrm{skill}}(o)
\right)
e^{-\lambda_c C_{\mathrm{conflict}}(o)},
\label{eq:hierarchical_reward}
\end{equation}
where \(R_{\mathrm{task}}\) measures the correctness of the final
page-level judgment,
\(R_{\mathrm{skill}}\) measures the quality of skill-grounded diagnosis
and structured attribution, and
\(C_{\mathrm{conflict}}\) penalizes unsupported evidence and
inconsistencies between local diagnoses and the final decision.

Following DAPO, rewards within each sampled group are converted into
group-relative advantages for policy optimization.
We adopt its clip-higher strategy, which uses asymmetric
clipping with a larger upper bound to preserve useful exploration, and dynamic sampling, which filters or resamples rollout groups
with insufficient reward variation to maintain informative learning
signals.
These mechanisms improve the stability and efficiency of
reinforcement learning while avoiding premature policy collapse.

\subsection{Self-evolving Skill Bank}
\label{sec:self_evolution}

The above training procedure produces the initial model before
deployment, corresponding to evolution round \(t=0\).
However, newly observed traffic and reviewed production feedback may reveal
evaluation patterns that are not fully covered by the initial
evaluation knowledge.
We address this adaptation problem through the
self-evolving skill bank.
SEEK enables lightweight knowledge adaptation by evolving the external
skill bank without immediately updating the router or evaluator.
During skill bank evolution, model checkpoints and routing
descriptions \(r_i^{(t)}\) remain fixed, while only the operational
guidance \(g_i^{(t)}\) is eligible for revision.

\subsubsection{Failure Attribution}

A disagreement between the SEEK prediction \(\widehat{y}\) and the
human reference \(y^{*}\) is treated as a candidate evolution signal.
A strong LLM serves as the \emph{evolution controller} and jointly
examines the query--list pair \((q,L)\), the selected skill set
\(\mathcal{S}_{x}^{(t)}\), current skill guidance, model prediction,
and human reference.

We categorize failures into three types.
A \emph{routing error} occurs when the required knowledge already
exists in the skill bank but the corresponding skill is not activated.
A \emph{skill knowledge error} occurs when the appropriate skill is
selected but its operational guidance is incomplete or outdated.
An \emph{execution error} occurs when both routing and guidance are
adequate, but the evaluator fails to apply them correctly.
Only skill knowledge errors enter the fast evolution loop;
routing and execution errors are accumulated for subsequent router or
evaluator updates.

\subsubsection{Recurrent Knowledge Revision}

To avoid overfitting to isolated failures, SEEK revises a skill only
when the same underlying knowledge gap recurs across multiple reviewed
samples.
For each skill \(S_i^{(t)}\), the evolution controller groups knowledge
errors that reflect the same missing evaluation principle into a
recurrent error cluster \(\mathcal{B}_{i}^{(t)}\).
It then performs a localized revision of the corresponding operational
guidance:
\begin{equation}
\widetilde{g}_{i}^{(t+1)}
=
\operatorname{Revise}
\left(
g_i^{(t)},
\mathcal{B}_{i}^{(t)}
\right),
\qquad
\widetilde{S}_{i}^{(t+1)}
=
\left(
r_i^{(t)},
\widetilde{g}_{i}^{(t+1)}
\right).
\label{eq:skill_revision}
\end{equation}

The revision is intentionally localized.
It may refine a decision boundary, evidence requirement, exception, or
previously missing evaluation pattern, while preserving the routing
semantics \(r_i^{(t)}\) and unrelated portions of the existing
guidance.
This allows new knowledge to be incorporated without unnecessarily
changing when the skill is activated.

\subsubsection{Replay-Gated Deployment}

Candidate revisions are validated before being deployed.
For each revised skill, we evaluate the candidate on a held-out
adaptation set \(\mathcal{D}_{i}^{\mathrm{new}}\), which contains
examples of the emerging pattern, and on a frozen historical replay set
\(\mathcal{D}^{\mathrm{replay}}\).
Both sets are associated with trusted human annotations.

Let \(M(\mathcal{D};S_i)\) denote the evaluation performance obtained
when skill \(S_i\) is used while all other system components remain
fixed.
A candidate revision is accepted only if it improves performance on the
newly observed pattern while preserving historical performance within a
tolerance \(\epsilon\):
\begin{equation}
\operatorname{Accept}
\left(
\widetilde{S}_{i}^{(t+1)}
\right)
=
\mathbb{I}
\left[
\begin{aligned}
M\left(
\mathcal{D}_{i}^{\mathrm{new}};
\widetilde{S}_{i}^{(t+1)}
\right)
&>
M\left(
\mathcal{D}_{i}^{\mathrm{new}};
S_i^{(t)}
\right),\\
M\left(
\mathcal{D}^{\mathrm{replay}};
\widetilde{S}_{i}^{(t+1)}
\right)
&\ge
M\left(
\mathcal{D}^{\mathrm{replay}};
S_i^{(t)}
\right)-\epsilon
\end{aligned}
\right].
\label{eq:replay_gate}
\end{equation}

Accepted candidates replace the previous skill version; otherwise, the
existing skill is retained.
In this way, SEEK separates adaptation across two timescales:
explicit evaluation knowledge can be updated rapidly through the skill
bank, while accumulated routing and execution errors are consolidated
through slower periodic model updates.

\section{Experiments}
\label{sec:experiments}

We conduct extensive experiments to evaluate SEEK from four
complementary perspectives.
First, we compare its listwise evaluation performance against
open-source LLMs, API-based LLMs, and
task-specific post-trained LLMs.
Second, we study the contribution of each major component, including
dynamic skill routing, evaluator post-training, and hierarchical
reward optimization.
Third, we investigate the continual-adaptation capability of the
external skill bank under frozen model parameters.
Finally, we evaluate the practical effectiveness and efficiency of
SEEK in a real online production environment.

\begin{table*}[t]
\centering
\small
\caption{
Overall performance on listwise search quality evaluation.
We report recall over five attribution dimensions, binary page-level
evaluation, and three-class quality evaluation.
Best results are shown in \textbf{bold}, and second-best results are
\underline{underlined}.
}
\label{tab:overall_main}
\vspace{-0.2cm}
\setlength{\tabcolsep}{2.2pt}
\renewcommand{\arraystretch}{1.08}
\begin{tabular}{
l
ccccc|
cc|
ccccc
}
\toprule
\multirow{2}{*}{\textbf{Method}}
& \multicolumn{5}{c|}{\textbf{Attribution Recall}}
& \multicolumn{2}{c|}{\textbf{Binary}}
& \multicolumn{5}{c}{\textbf{Three-Class}} \\
\cmidrule(lr){2-6}
\cmidrule(lr){7-8}
\cmidrule(lr){9-13}
& \textbf{Rel.}
& \textbf{Qual.}
& \textbf{Div.}
& \textbf{Auth.}
& \textbf{Het.}
& \textbf{Macro-F1}
& \textbf{Acc.}
& \textbf{Good F1}
& \textbf{Fair F1}
& \textbf{Bad F1}
& \textbf{Macro-F1}
& \textbf{Acc.} \\
\midrule
\multicolumn{13}{l}{\textit{Open-source LLMs}} \\
Qwen3-8B~\cite{yang2025qwen3}
& 0.5566 & 0.5585 & 0.1624 & 0.2272 & 0.1046
& 0.4345 & 0.4894
& 0.3186 & 0.4217 & 0.4562 & 0.3988 & 0.4085 \\
DeepSeek-R1-Distill-Qwen-7B~\cite{guo2025deepseek}
& 0.6847 & 0.6416 & 0.3185 & 0.3027 & 0.1942
& 0.5888 & 0.5968
& 0.3810 & 0.4676 & 0.5268 & 0.4585 & 0.4706 \\
Gemma-3-12B-IT~\cite{team2025gemma}
& 0.7168 & 0.6635 & 0.3642 & 0.3298 & 0.2187
& 0.6125 & 0.6216
& 0.3955 & 0.4781 & 0.5434 & 0.4723 & 0.4859 \\
\midrule
\multicolumn{13}{l}{\textit{API-based LLMs}} \\
Qwen3-235B-A22B~\cite{yang2025qwen3}
& 0.8078 & 0.6498 & 0.4450 & 0.4076 & 0.2832
& 0.6478 & 0.6676
& 0.4247 & 0.5046 & 0.5639 & 0.4977 & 0.5114 \\
MiniMax-M2.5~\cite{minimax_m25}
& 0.8256 & 0.6365 & 0.4413 & 0.3984 & 0.2913
& 0.6700 & 0.6866
& 0.4307 & 0.4989 & 0.5959 & 0.5085 & 0.5227 \\
Gemini 3.1 Pro~\cite{gemini31pro}
& 0.8396 & \underline{0.8114} & 0.5883 & 0.4897 & 0.3985
& 0.7241 & 0.7364
& 0.6198 & 0.5721 & 0.6804 & 0.6241 & 0.6326 \\
GPT-5.6 Sol~\cite{gpt56}
& \underline{0.8518} & 0.8037 & 0.6065 & \underline{0.5078} & \underline{0.4146}
& \underline{0.7358} & \underline{0.7451}
& 0.6427 & \underline{0.5776} & \underline{0.6932} & 0.6378 & 0.6469 \\

\midrule
\multicolumn{13}{l}{\textit{Task-specific Post-trained LLMs}} \\

Qwen3-8B (SFT+DPO)~\cite{tang2025lref,lu2025vlm}
& 0.8015 & 0.7426 & 0.5638 & 0.3725 & 0.3261
& 0.6876 & 0.7068
& 0.4685 & 0.5281 & 0.6124 & 0.5363 & 0.5613 \\

Qwen3-8B (SFT+GRPO)~\cite{lu2025lore,zeng2026optimizing,yang2025taosr}
& 0.8462 & \textbf{0.8215} & \underline{0.6189} & 0.4381 & 0.3584
& 0.7206 & 0.7386
& \underline{0.6513} & \textbf{0.6107} & 0.6634 & \underline{0.6418} & \underline{0.6527} \\

\midrule

\textbf{Qwen3-8B (SEEK)}
& \textbf{0.8621} & 0.8102 & \textbf{0.6210} & \textbf{0.5240} & \textbf{0.4305}
& \textbf{0.7374} & \textbf{0.7520}
& \textbf{0.6715} & 0.5586 & \textbf{0.7364} & \textbf{0.6555} & \textbf{0.6618} \\

\bottomrule
\end{tabular}
\end{table*}

\subsection{Experimental Setup}
\label{sec:exp_settings}

\subsubsection{Dataset and Annotation}

We construct the dataset from query--list pairs sampled from
Kuaishou search logs.
Each retrieved result contains information including the title,
image caption, OCR and ASR text, creator information, interaction
statistics, and ranking position.

Each query--list pair is annotated with a page-level quality label from
\{\textit{good}, \textit{fair}, \textit{bad}\}.
For \textit{fair} and \textit{bad} pages, annotators additionally
provide attribution diagnoses over five dimensions:
\textit{Relevance}, \textit{Quality}, \textit{Diversity},
\textit{Authority}, and \textit{Heterogeneous}.
A fair or bad page may correspond to multiple dimensions.
Annotation disagreements are resolved by experienced search assessors.

After cleaning and deduplication, we obtain 169,434 training
query--list pairs and an independent held-out test set of 17,000
pairs.
We apply stratified, label-aware filtering when constructing
the held-out set.
This procedure ensures sufficient representation of all three
page-level quality labels and each of the five attribution dimensions.
Detailed dataset statistics are reported in
Appendix~\ref{app:annotation}.

\subsubsection{Baselines}

We compare SEEK with three complementary groups of baselines.

\paragraph{Open-source LLMs.}
We include Qwen3-8B~\cite{yang2025qwen3},
DeepSeek-R1-Distill-Qwen-7B~\cite{guo2025deepseek},
and Gemma-3-12B-IT~\cite{team2025gemma}.
These models represent compact general-purpose
LLMs without task-specific adaptation to our listwise search evaluation task.

\paragraph{API-based LLMs.}
We further compare with Qwen3-235B-A22B~\cite{yang2025qwen3},
MiniMax-M2.5~\cite{minimax_m25},
Gemini 3.1 Pro~\cite{gemini31pro},
and GPT-5.6 Sol~\cite{gpt56}.
These models provide strong general-purpose judging capabilities at
substantially larger scales.
All models receive the same query--list input,
evaluation criteria, and output specification.

\paragraph{Task-specific Post-trained LLMs.}
We additionally compare with representative parameter-level tuning using the same Qwen3-8B backbone.
Direct Preference Optimization (DPO)~\cite{rafailov2023direct}
optimizes the model directly from preference supervision without
training a separate reward model, whereas Group Relative Policy
Optimization (GRPO)~\cite{shao2024deepseekmath}
performs reinforcement learning with group-relative advantages
computed from multiple sampled responses.

Following prior search evaluation and relevance modeling methods,
we instantiate two task-adapted baselines:
SFT followed by DPO~\cite{tang2025lref,lu2025vlm},
and SFT followed by GRPO
~\cite{lu2025lore,zeng2026optimizing,yang2025taosr}.
These baselines internalize task-specific evaluation knowledge through
supervised fine-tuning followed by preference
optimization or reinforcement learning, providing controlled
comparisons with SEEK's externalized, skill-grounded knowledge design.

\subsubsection{Evaluation Metrics}

We evaluate page-level prediction under both binary and three-class
settings.
For binary evaluation, \textit{good} pages are treated as the positive
class, while \textit{fair} and \textit{bad} pages are merged into the
negative class.
We report Macro-F1 and Accuracy, where Macro-F1 is computed as the
unweighted average of the F1.

For three-class evaluation, we retain the original
\{\textit{good}, \textit{fair}, \textit{bad}\} label space and report
class-wise F1, Macro-F1, and overall Accuracy.
To assess fine-grained attribution quality, we report Recall over the
five attribution dimensions:
\textit{Relevance}, \textit{Quality}, \textit{Diversity},
\textit{Authority}, and \textit{Heterogeneous}.

For continual adaptation, we additionally report Binary F1 and
Attribution F1 to measure adaptation effectiveness and the quality of
fine-grained diagnosis under evolving evaluation criteria.

\subsubsection{Implementation Details}

The listwise evaluator is initialized from Qwen3-8B.
All offline training experiments are conducted on NVIDIA H800 GPUs,
while production inference is deployed on NVIDIA L20 GPU clusters.
Additional implementation and optimization details are
provided in Appendix~\ref{app:implementation}.

\subsection{Overall Performance}
\label{sec:offline_eval}

Table~\ref{tab:overall_main} summarizes the overall results on
listwise search quality evaluation.
SEEK consistently outperforms its Qwen3-8B backbone across page-level
prediction and fine-grained attribution, with substantial gains under
both binary and three-class settings.
This indicates that SEEK improves not only the detection of problematic
result pages, but also the discrimination of more subtle quality
differences among \textit{good}, \textit{fair}, and \textit{bad}
search experiences.

Task-specific post-training substantially narrows the gap, making
SFT+GRPO a strong same-backbone baseline.
Nevertheless, SEEK further improves three-class Macro-F1 from 0.6418
to 0.6555 while also achieving better binary evaluation performance.
This result suggests that externalizing
evaluation knowledge and dynamically selecting relevant criteria
provides complementary benefits beyond parameter-level adaptation.

The advantage is particularly evident in fine-grained attribution.
SEEK achieves the best results on four of the five dimensions, with
the largest improvements over SFT+GRPO appearing on
\textit{Authority} and \textit{Heterogeneous}.
These dimensions depend more heavily on task-specific evaluation
criteria and cross-result interactions, which are difficult to capture
through a single static judging policy.
The results therefore support the benefit of dynamically selecting and
providing explicit skill guidance for each query--list pair.

\subsection{Ablation Study}
\label{sec:component_analysis}

We conduct ablation studies to examine four key aspects of SEEK:
the complementary contributions of the skill router and listwise
evaluator, the effect of evaluator post-training, the hierarchical
reward design, and the effectiveness of dynamic skill routing.

\subsubsection{Skill Router and Evaluator}

We disentangle the effects of dynamic skill routing and evaluator
post-training in Table~\ref{tab:ablation}.
When routing is disabled, the evaluator receives the complete Skill
Bank; when evaluator training is disabled, the original Qwen3-8B
backbone is used.

Dynamic routing consistently improves over exposing all skills,
showing that instance-specific skill selection is preferable to
indiscriminate knowledge injection.
Evaluator post-training provides a further substantial gain by
learning task-specific listwise evaluation capabilities.
Combining both components yields the best performance, confirming
their complementary roles: the evaluator learns how to perform the
task, while the router determines which external knowledge should be
applied to each query--list pair.


\begin{table}[t]
\centering
\small
\caption{
Ablation of dynamic skill routing and evaluator post-training.
\textit{All Skills} provides the complete skill bank to every
query--list pair without routing, whereas \textit{Dynamic} selects an
instance-specific subset using the skill router.
\textit{Base} denotes the original Qwen3-8B without task-specific
post-training, and \textit{Trained} denotes the trained
listwise evaluator.
}
\label{tab:ablation}
\vspace{-0.2cm}

\setlength{\tabcolsep}{2.2pt}
\renewcommand{\arraystretch}{1.08}

\begin{tabular}{ll|ccccc|c}
\toprule
\multicolumn{2}{c|}{\textbf{Configuration}}
& \multicolumn{5}{c|}{\textbf{Attribution Recall}}
& \multirow{2}{*}{\textbf{Acc.}} \\
\cmidrule(lr){1-2}
\cmidrule(lr){3-7}
\textbf{Routing}
& \textbf{Evaluator}
& \textbf{Rel.}
& \textbf{Qual.}
& \textbf{Div.}
& \textbf{Auth.}
& \textbf{Het.}
& \\
\midrule

All Skills
& Base
& 0.5566
& 0.5585
& 0.1624
& 0.2272
& 0.1046
& 0.4894 \\

Dynamic
& Base
& 0.7842
& 0.7146
& 0.4218
& 0.3518
& 0.2765
& 0.6485 \\

All Skills
& Trained
& \underline{0.8284}
& \underline{0.7626}
& \underline{0.4987}
& \underline{0.4124}
& \underline{0.3363}
& \underline{0.6913} \\

Dynamic
& Trained
& \textbf{0.8621}
& \textbf{0.8102}
& \textbf{0.6210}
& \textbf{0.5240}
& \textbf{0.4305}
& \textbf{0.7520} \\

\bottomrule
\end{tabular}
\end{table}

\subsubsection{Evaluator Optimization}

We examine the effect of evaluator post-training in
Table~\ref{tab:evaluator_ablation}, comparing the original Qwen3-8B
backbone, SFT alone, SFT followed by GRPO, and SFT followed by DAPO.
SFT yields a substantial improvement over the base model, showing that
teacher-generated structured trajectories provide effective supervision
for learning listwise evaluation behavior.
Subsequent reinforcement learning further improves both page-level
prediction and fine-grained attribution, with DAPO achieving the best
overall performance.
These results suggest that SFT establishes the task-specific evaluation
capability, while reinforcement learning further aligns the
skill-conditioned evaluation process with the final page-level
judgment.


\begin{table}[t]
\centering
\small
\caption{
Ablation of evaluator optimization. ``+'' indicates that the
corresponding training stage is enabled.
}
\label{tab:evaluator_ablation}
\vspace{-0.2cm}
\setlength{\tabcolsep}{2.2pt}
\renewcommand{\arraystretch}{1.08}
\begin{tabular}{ccc|ccccc|c}
\toprule
\multicolumn{3}{c|}{\textbf{Evaluator}}
& \multicolumn{5}{c|}{\textbf{Attribution Recall}}
& \multirow{2}{*}{\textbf{Acc.}} \\
\cmidrule(lr){1-3}
\cmidrule(lr){4-8}
\textbf{SFT} & \textbf{GRPO} & \textbf{DAPO}
& \textbf{Rel.} & \textbf{Qual.} & \textbf{Div.} & \textbf{Auth.} & \textbf{Het.} & \\
\midrule
-- & -- & -- & 0.5566 & 0.5585 & 0.1624 & 0.2272 & 0.1046 & 0.4894 \\
+ & -- & -- & 0.8332 & 0.7886 & 0.5158 & 0.4533 & 0.3800 & 0.6251 \\
+ & + & -- & \underline{0.8547} & \underline{0.8016} & \underline{0.5865} & \underline{0.4967} & \underline{0.4127} & \underline{0.7385} \\
\textbf{+} & -- & \textbf{+} & \textbf{0.8621} & \textbf{0.8102} & \textbf{0.6210} & \textbf{0.5240} & \textbf{0.4305} & \textbf{0.7520} \\
\bottomrule
\end{tabular}
\end{table}

\subsubsection{Hierarchical Reward}

We examine the hierarchical reward design in
Table~\ref{tab:reward_ablation} by progressively adding
skill-level supervision and the conflict penalty to the task-level
reward.

Using \(R_{\mathrm{task}}\) alone already provides strong page-level
supervision, but yields weaker fine-grained attribution.
Adding \(R_{\mathrm{skill}}\) consistently improves diagnostic quality,
while \(C_{\mathrm{conflict}}\) further encourages consistency between
skill-level diagnoses and the final judgment.
The full objective performs best overall, indicating that reliable
listwise evaluation benefits from jointly optimizing prediction
correctness, skill-grounded diagnosis, and cross-level consistency.

\begin{table}[t]
\centering
\small
\caption{
Ablation of the hierarchical reward in SEEK.
``+'' indicates that the corresponding reward component is enabled.
}
\label{tab:reward_ablation}
\vspace{-0.2cm}
\setlength{\tabcolsep}{2.5pt}
\renewcommand{\arraystretch}{1.08}

\begin{tabular}{ccc|ccccc|c}
\toprule
\multicolumn{3}{c|}{\textbf{Reward}}
& \multicolumn{5}{c|}{\textbf{Attribution Recall}}
& \multirow{2}{*}{\textbf{Acc.}} \\
\cmidrule(lr){1-3}
\cmidrule(lr){4-8}
\textbf{$R_{\text{task}}$}
& \textbf{$R_{\text{skill}}$}
& \textbf{$C_{\text{conflict}}$}
& \textbf{Rel.}
& \textbf{Qual.}
& \textbf{Div.}
& \textbf{Auth.}
& \textbf{Het.}
& \\
\midrule

+
& --
& --
& 0.8502 & 0.7950 & 0.5581 & 0.4746 & 0.3964 & 0.7126 \\

+
& +
& --
& \underline{0.8575}
& \underline{0.8027}
& \underline{0.5941}
& \underline{0.5018}
& \underline{0.4160}
& \underline{0.7382} \\

\textbf{+}
& \textbf{+}
& \textbf{+}
& \textbf{0.8621}
& \textbf{0.8102}
& \textbf{0.6210}
& \textbf{0.5240}
& \textbf{0.4305}
& \textbf{0.7520} \\

\bottomrule
\end{tabular}
\end{table}

\subsubsection{Dynamic Skill Routing}
\label{sec:routing_analysis}

We compare three skill-access strategies in
Table~\ref{tab:routing_analysis}.
Dynamic Routing substantially outperforms indiscriminate access to the
entire skill bank, while requiring only a small fraction of the
available skills for each instance.
This indicates that instance-specific selection
reduces unnecessary context and potential criterion interference.
The remaining gap to Oracle Routing suggests that better skill
selection could provide further gains.


\begin{table}[t]
\centering
\small
\caption{
Comparison of skill-access strategies.
\textit{Oracle Routing} uses reference skill
annotations as an upper bound.
}
\label{tab:routing_analysis}
\vspace{-0.2cm}

\setlength{\tabcolsep}{5.0pt}
\renewcommand{\arraystretch}{1.08}

\begin{tabular}{lccc}
\toprule
\multirow{2}{*}{\textbf{Skill Access}}
& \multicolumn{2}{c}{\textbf{Evaluation Quality}}
& \multirow{2}{*}{\textbf{Avg. Skills}} \\
\cmidrule(lr){2-3}
& \textbf{Acc.}
& \textbf{Attr. F1}
& \\
\midrule

All Skills
& 0.691
& 0.664
& 12.0 \\

Dynamic Routing
& \underline{0.752}
& \underline{0.743}
& 3.4 \\

Oracle Routing
& \textbf{0.768}
& \textbf{0.758}
& 3.1 \\

\bottomrule
\end{tabular}
\end{table}

\subsection{Continual Adaptation}
\label{sec:self_evolution_exp}

We evaluate whether SEEK can adapt to emerging evaluation criteria by
updating the external skill bank.
The experiments are conducted on a newly collected post-deployment
test set whose distribution differs from the held-out benchmark used
for offline evaluation.
This dataset better reflects the evolving query patterns and diagnostic dimensions encountered in real
production environments.
We examine both adaptation to newly observed patterns and stability on
historical data.

\subsubsection{Knowledge and Parameter Adaptation}

\begin{figure}[t]
    \centering
    \includegraphics[width=\columnwidth]
    {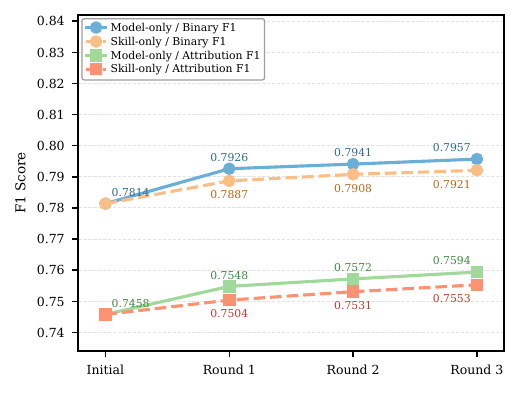}
    \caption{
    Multi-round comparison of parameter-based adaptation and
    Skill-only Evolution.
    Both strategies are evaluated using Binary F1 and Attribution F1
    under the same reviewed evolution data.
    }
    \label{fig:multi_round_adaptation}
    \vspace{-0.2cm}
\end{figure}

Figure~\ref{fig:multi_round_adaptation} compares two strategies under
the same sequence of reviewed production feedback.
\emph{Model-only Update} fine-tunes the evaluator while
keeping the skill bank fixed, whereas \emph{Skill-only Evolution}
updates only the skill bank with both model components frozen.

Both strategies improve consistently as reviewed samples are
incorporated.
After three rounds, Model-only Update increases Binary F1 from 0.7814
to 0.7957 and Attribution F1 from 0.7458 to 0.7594.
Without modifying either model component, Skill-only Evolution reaches
0.7921 Binary F1 and 0.7553 Attribution F1.
The remaining gaps to parameter-based adaptation are only 0.0036 and
0.0041, respectively, showing that a substantial fraction of the
adaptation benefit can be recovered by updating explicit evaluation
knowledge alone.

The largest gains occur in the first evolution round, followed by
progressively smaller improvements in subsequent rounds.
This trend is consistent with recurrent and reusable knowledge gaps
being absorbed early, while later rounds mainly capture more
specialized patterns.
Overall, the capability--knowledge separation underlying SEEK provides a lightweight fast path for incorporating new
evaluation knowledge without frequent model retraining.

\subsubsection{Replay Stability}

We further evaluate backward stability on a frozen historical replay
set after the same three adaptation rounds.
As shown in Table~\ref{tab:replay_stability}, Skill-only Evolution
largely preserves historical performance, with only
0.0008 Binary-F1 forgetting and
0.0005 Attribution-F1 forgetting.
In contrast, Model-only Update causes noticeably larger degradation,
suggesting that parameter adaptation introduces greater interference
with previously learned evaluation behavior.

Together with the adaptation results above, this demonstrates the
advantage of SEEK's design:
Skill-only Evolution provides a fast and backward-stable path for
absorbing new evaluation knowledge, while parameter updates can be less frequent.

\begin{table}[t]
\centering
\small
\caption{
Historical replay stability after three adaptation rounds.
Both strategies use the same reviewed feedback sequence and frozen
replay set.
Forgetting is the F1 decrease relative to the initial system.
}
\label{tab:replay_stability}
\vspace{-0.2cm}
\setlength{\tabcolsep}{3.8pt}
\renewcommand{\arraystretch}{1.08}

\begin{tabular}{lcccc}
\toprule
\textbf{Strategy}
& \textbf{Binary F1}
& \textbf{Forget.}
& \textbf{Attr. F1}
& \textbf{Forget.} \\
\midrule

Initial System
& 0.7862
& --
& 0.7481
& -- \\

Model-only Update
& 0.7786
& 0.0076
& 0.7397
& 0.0084 \\

Skill-only Evolution
& 0.7854
& 0.0008
& 0.7476
& 0.0005 \\

\bottomrule
\end{tabular}
\end{table}

\subsection{Production Evaluation}
\label{sec:production_eval} 

Finally, we evaluate SEEK in a real-world production
search evaluation workflow.
Table~\ref{tab:production_evaluation} compares SEEK with outsourced
human review and multi-round human assessment on 24,000 query--list
pairs, using expert-calibrated judgments as the reference labels.

SEEK achieves a favorable balance between judgment quality and
efficiency.
It reaches an Accuracy of 0.842, outperforming both outsourced review
(0.783) and multi-round assessment (0.830), while completing the
entire workload in only 0.9 hours.
Compared with outsourced review, this corresponds to a
40.6$\times$ speedup.
Notably, although multi-round assessment improves human review quality,
it incurs substantially higher time cost, whereas SEEK attains stronger
agreement with expert-calibrated labels at only a fraction of the
evaluation time.

Beyond page-level predictions, SEEK also produces structured
attribution that identify the quality dimensions underlying each
judgment.
These diagnostic signals can directly support large-scale quality
monitoring, failure analysis, and downstream search-system iteration.
Therefore, SEEK can serve as a scalable
and practical evaluation component for continuous industrial search
development.


\begin{table}[t]
\centering
\small
\caption{
Production evaluation on 24,000 online query--list pairs.
Accuracy is measured against expert-calibrated reference labels, and
speedup is computed relative to outsourced human review.
}
\label{tab:production_evaluation}
\vspace{-0.2cm}

\setlength{\tabcolsep}{4.0pt}
\renewcommand{\arraystretch}{1.08}

\begin{tabular}{lrrr}
\toprule
\textbf{Evaluator}
& \textbf{Time} $\downarrow$
& \textbf{Speedup} $\uparrow$
& \textbf{Accuracy} $\uparrow$ \\
\midrule

Outsourced Review
& 36.5 h
& 1.0$\times$
& 0.783 \\

Multi-round Review
& 58.2 h
& 0.6$\times$
& 0.830 \\

SEEK
& \textbf{0.9 h}
& \textbf{40.6$\times$}
& \textbf{0.842} \\

\bottomrule
\end{tabular}
\end{table}

\section{Conclusion}

In this paper, we propose SEEK, a skill-grounded framework for
industrial listwise search evaluation that decouples evolving
domain knowledge from relatively stable model capabilities.
SEEK externalizes heterogeneous evaluation criteria into a modular
skill bank, dynamically routes task-relevant skills for each
query--list pair, and employs a task-adapted listwise evaluator to
produce page-level judgments and structured attribution.
A human-aligned training pipeline specializes the skill router and
evaluator, while the self-evolving skill bank enables rapid knowledge
adaptation without requiring immediate model retraining.

Extensive offline, continual-adaptation, and production experiments
show that SEEK improves listwise judgment and fine-grained diagnosis,
supports effective adaptation through external knowledge updates with
frozen model parameters, and substantially improves evaluation
efficiency in production.
Overall, the results demonstrate the value of separating evolvable
evaluation knowledge from model reasoning capabilities, providing a
practical path toward more scalable and adaptable industrial search
evaluation.

\clearpage
\bibliographystyle{ACM-Reference-Format}
\bibliography{reference}


\appendix

\section{Skill Bank Details}
\label{app:skill_bank}

SEEK externalizes task-specific evaluation knowledge into a modular
skill bank.
In our implementation, the skill bank contains 12 diagnostic skills
covering five major dimensions of industrial search quality:
\textit{Relevance}, \textit{Quality}, \textit{Diversity},
\textit{Authority}, and \textit{Heterogeneous}.
Rather than treating each dimension as a single coarse criterion, we
decompose it into more specific skills that capture recurring
evaluation principles and decision boundaries.
Table~\ref{tab:skill_taxonomy} summarizes the skill taxonomy used in
our experiments.

\begin{table*}[t]
\centering
\small
\caption{
Taxonomy of the 12 diagnostic skills used in SEEK.
Each skill captures a reusable evaluation criterion within one
high-level search-quality dimension.
}
\label{tab:skill_taxonomy}
\vspace{-0.1cm}
\setlength{\tabcolsep}{5pt}
\renewcommand{\arraystretch}{1.08}

\begin{tabular}{
p{0.15\textwidth}
p{0.22\textwidth}
p{0.55\textwidth}
}
\toprule
\textbf{Dimension}
& \textbf{Skill}
& \textbf{Evaluation Focus} \\
\midrule

\multirow{3}{*}{Relevance}
& MainIntentMatch
& Whether retrieved content matches the dominant intent expressed by
the query. \\

& TaskNeedSatisfaction
& Whether the result page satisfies the underlying user need rather
than merely exhibiting lexical or topical overlap. \\

& RankingFaithfulness
& Whether highly relevant and useful results are placed at appropriate
positions relative to weaker results. \\

\midrule

\multirow{3}{*}{Quality}
& InformationValueCheck
& Whether retrieved results provide substantive and useful information
rather than low-value, vague, or weakly informative content. \\

& FactualReliabilityCheck
& Whether factual or professional claims are sufficiently supported by
reliable evidence and avoid unsupported assertions. \\

& RiskComplianceCheck
& Whether risky, misleading, or otherwise problematic content
introduces quality or compliance concerns. \\

\midrule

\multirow{3}{*}{Diversity}
& SourceDiversityCheck
& Whether the result page avoids excessive concentration on the same
or highly similar content sources. \\

& TemplateDuplication
& Whether multiple results exhibit substantial semantic, structural,
or template-level redundancy. \\

& SubIntentCoverageCheck
& Whether distinct and plausible sub-intents of the query receive
adequate coverage across the result page. \\

\midrule

\multirow{2}{*}{Authority}
& SourceAuthorityCheck
& Whether source credibility and authority match the level of
expertise required by the query. \\

& EvidenceProfessionalism
& Whether professional or high-confidence information is supported by
appropriate credentials, evidence, or domain expertise. \\

\midrule

Heterogeneous
& HeterogeneousFulfillment
& Whether heterogeneous result types and information sources
complement one another in satisfying complex user needs. \\

\bottomrule
\end{tabular}
\end{table*}

\paragraph{Skill Representation.}
As introduced in Section~\ref{sec:framework}, each skill is represented
as \(S_i=(r_i,g_i)\), where \(r_i\) is a compact routing description
and \(g_i\) is the corresponding operational guidance.
The two components provide different views of the same evaluation
criterion.
The routing description specifies \emph{when} a skill is applicable
and is used by the skill router for instance-specific selection.
In contrast, the operational guidance specifies \emph{how} the
criterion should be applied and is exposed to the listwise evaluator
only after the skill is activated.

Operational guidance is structured around five types of information:
the evaluation objective, required evidence, decision rules, boundary
conditions, and attribution requirements.
The objective defines the quality property being assessed;
evidence requirements specify which signals in the query--list pair
should support the judgment;
decision rules describe how the evidence should be interpreted;
boundary conditions prevent the criterion from being applied outside
its intended scope; and attribution requirements specify how detected
problems should be connected to the final page-level diagnosis.
This representation keeps routing descriptions compact while allowing
the evaluator to access sufficiently detailed and actionable knowledge.

\paragraph{Example Skill: SourceAuthorityCheck.}
As a representative example, the routing description of
\textit{SourceAuthorityCheck} activates the skill when satisfying the
query requires professional, authoritative, or high-confidence
information.
Its operational guidance asks the evaluator to determine whether the
credibility of retrieved sources matches the level of expertise
required by the user intent.
Relevant evidence includes creator identity, professional credentials,
institutional affiliation, source reputation, supporting evidence, and
ranking position.
For professional or high-stakes informational needs, stronger source
credibility is required, and popularity or interaction statistics alone
should not be treated as sufficient evidence of authority.
Conversely, professional authority should not be imposed on ordinary
entertainment or personal-experience queries unless the query
explicitly requires such expertise.
When an authority issue is identified, the evaluator attributes the
problem to the affected results and explains its impact on the overall
quality of the result page.

This decomposition also allows multiple skills to be activated for the
same query--list pair.
For example, a result containing an unsupported professional claim may
simultaneously require relevance checking, factual-reliability
assessment, and source-authority verification.
The multi-label router therefore selects a compact combination of
complementary skills rather than assigning each instance to a single
evaluation category.

\section{Implementation Details}
\label{app:implementation}

\paragraph{Model Configuration.}
SEEK uses Qwen3-0.6B as the skill router and Qwen3-8B as the
listwise evaluator.
Qwen3-235B-A22B is used as the teacher model for routing-target
generation, structured trajectory construction, and skill evolution.
The evaluator supports a maximum input length of 16,384 tokens and a
maximum output length of 4,096 tokens.
Each query--list instance includes title, image caption, OCR/ASR text,
creator information, interaction statistics, and ranking position.

\paragraph{Skill Router.}
The router is trained as a 12-way multi-label classifier using
teacher-derived routing targets.
We optimize it for three epochs with AdamW, a learning rate of
\(1\times10^{-5}\), weight decay of \(0.01\), and an effective batch
size of 32.
The activation threshold is set to \(\tau=0.5\); if no skill exceeds
the threshold, the highest-scoring skill is retained.
This setting activates approximately three to four skills per
query--list pair.

\paragraph{Evaluator Training.}
The evaluator is first trained with SFT for two epochs using a learning
rate of \(2\times10^{-5}\), batch size 4, and an effective batch size
of 32.
We then apply DAPO with a learning rate of \(1\times10^{-6}\) and a
group size of 6.
Rollouts use temperature \(0.8\) and top-\(p=0.95\).
For the hierarchical reward, we set
\(\lambda_s=0.5\) and \(\lambda_c=1.0\).
All training uses BF16 precision, and loss is applied only to
assistant-generated tokens.

\paragraph{Skill Evolution and Inference.}
During skill-only evolution, the router and evaluator remain frozen and
only the operational guidance \(g_i\) is revised.
A knowledge gap is considered recurrent when it appears in at least
three reviewed samples.
Candidate revisions are validated on both emerging cases and a frozen
replay set of 5,000 historical query--list pairs, with replay tolerance
\(\epsilon=0.002\).
Offline training is conducted on NVIDIA H800 GPUs, while production
inference runs on NVIDIA L20 GPU clusters.
The final evaluator uses deterministic decoding.

\section{Qualitative Analysis}
\label{app:qualitative}

We provide a qualitative case study to illustrate how SEEK converts
recurrent production failures into reusable evaluation knowledge.
Rather than storing individual failure cases as additional
demonstrations, the self-evolving skill bank identifies their shared
evaluation principle and performs a localized revision of the
corresponding operational guidance.

\paragraph{Representative Skill Evolution Case.}
We consider a recurrent pattern involving health-related search
results.
In several reviewed query--list pairs, the evaluator correctly
activated skills related to factual reliability and source authority,
but still assigned overly favorable judgments to results containing
strong health claims.
Inspection by the evolution controller attributed these failures to a
\emph{skill knowledge error}: the relevant skills were correctly
selected, while the existing operational guidance did not sufficiently
distinguish ordinary informational claims from high-stakes claims that
require stronger evidence and source credibility.

Table~\ref{tab:skill_evolution_case} summarizes the resulting
knowledge revision.
The original guidance was not replaced wholesale.
Instead, the Evolution Controller introduced a localized decision rule
requiring stronger evidential support for medical or other high-stakes
claims, while preserving the original routing semantics and unrelated
evaluation rules.

\begin{table}[t]
\centering
\small
\caption{
Representative localized revision during Skill-only Evolution.
}
\label{tab:skill_evolution_case}
\vspace{-0.1cm}
\setlength{\tabcolsep}{4pt}
\renewcommand{\arraystretch}{1.10}

\begin{tabular}{
p{0.32\columnwidth}
p{0.60\columnwidth}
}
\toprule
\textbf{Stage} & \textbf{Description} \\
\midrule

Recurrent Failure
&
Several reviewed results make strong health-related claims but provide
insufficient supporting evidence or source credibility, while the
evaluator judges them too favorably. \\

\midrule

Failure Attribution
&
The appropriate factual-reliability and authority skills are already
activated, indicating a skill knowledge error rather than a routing
error. \\

\midrule

Knowledge Gap
&
The existing guidance does not explicitly require stronger evidence
standards for medical or other high-stakes claims. \\

\midrule

Localized Revision
&
For high-stakes claims, require stronger supporting evidence and
credible sources; popularity or interaction signals alone should not
be treated as sufficient evidence of reliability or authority. \\

\midrule

Replay Validation
&
The candidate revision is evaluated on both newly reviewed cases and
the frozen historical replay set, and is deployed only if it improves
the emerging pattern without exceeding the allowed replay degradation. \\

\bottomrule
\end{tabular}
\end{table}

\paragraph{Analysis.}
This case highlights three properties of the evolution mechanism.
First, SEEK evolves \emph{evaluation knowledge} rather than merely
memorizing newly observed samples.
Multiple local failures are abstracted into a reusable decision rule
that can generalize to future query--list pairs exhibiting the same
underlying pattern.
Second, the update is localized to the operational guidance \(g_i\);
the routing description \(r_i\) remains unchanged because the original
skill activation behavior is already correct.
Finally, replay-gated validation constrains the revision from improving
newly emerging cases at the expense of previously stable traffic.

The example therefore illustrates the intended two-timescale
adaptation of SEEK.
Recurrent knowledge gaps can be addressed rapidly through explicit
skill bank revisions, whereas failures caused by incorrect routing or
improper skill execution are accumulated for slower periodic updates
of the router or evaluator.

\section{Annotation Protocol and Dataset Statistics}
\label{app:annotation}

Our annotation protocol follows the production search-quality
assessment process for short-video search.
Given a query \(q\) and its ranked result list
\(L=[d_1,\ldots,d_K]\), annotators evaluate the result page as a whole
rather than independently judging each retrieved item.
They jointly consider result-level evidence, ranking positions, and
cross-result interactions, such as redundancy, intent coverage, source
composition, and the overall usefulness of the result page.

\paragraph{Page-level Quality Labels.}
Each query--list pair is assigned one of three quality labels:
\textit{good}, \textit{fair}, or \textit{bad}.
A \textit{good} page adequately satisfies the dominant user intent and
contains no material defect that substantially affects the overall
search experience.
A \textit{fair} page remains generally useful but contains noticeable
quality issues, such as several weakly relevant results, moderate
redundancy, insufficient intent coverage, or localized content-quality
problems.
A \textit{bad} page contains severe or systematic defects that
substantially impair the overall search experience, such as major
intent mismatch, pervasive low-quality content, serious ranking
problems, or multiple interacting deficiencies.

Importantly, the final label is not determined solely by the number of
problematic results.
Annotators additionally consider the severity of each issue, its
ranking position, the fraction of the page affected, and whether
multiple problems jointly degrade the overall user experience.
The distinction between \textit{fair} and \textit{bad} therefore
reflects both the scope and the severity of the observed deficiencies.

\begin{table}[t]
\centering
\small
\caption{
Semantics of the page-level quality labels.
}
\label{tab:label_definition}
\vspace{-0.1cm}
\setlength{\tabcolsep}{4pt}
\renewcommand{\arraystretch}{1.06}

\begin{tabular}{lp{0.70\columnwidth}}
\toprule
\textbf{Label} & \textbf{Definition} \\
\midrule

Good
&
The result page adequately satisfies the search intent and contains no
material issue that substantially affects the overall experience. \\

Fair
&
The page is generally useful, but contains noticeable, localized, or
moderate quality problems. \\

Bad
&
Severe or systematic problems substantially degrade the overall search
experience. \\

\bottomrule
\end{tabular}
\end{table}

\paragraph{Annotation Procedure.}
Annotators are provided with the query, the complete ranked result
list, and the signals available to the production evaluator, including
titles, image captions, OCR and ASR text, creator information,
interaction statistics, and ranking positions.
They first infer the dominant user intent and inspect whether the
result page satisfies that intent.
They then examine individual results and cross-result interactions,
assign the page-level quality label, and, for \textit{fair} or
\textit{bad} pages, identify the dimensions responsible for the
degradation together with supporting evidence.

When multiple issues coexist, annotators are instructed to record all
major dimensions that materially contribute to the final judgment
rather than forcing each page into a single error category.
Ambiguous or conflicting cases are reviewed by experienced search
assessors, whose adjudicated decisions are used as the final reference
annotations for model training and evaluation.

\paragraph{Dataset Statistics.}

Table~\ref{tab:data_statistics} reports the detailed statistics of the
training and held-out evaluation sets.
The test set contains 17,000 query--list pairs and is used exclusively
for offline evaluation.
For binary evaluation, \textit{good} pages are treated as the positive
class, while \textit{fair} and \textit{bad} pages are merged into the
negative class.
Attribution is formulated as a multi-label diagnosis task; therefore,
a single page may be associated with multiple attribution dimensions.

\begin{table}[t]
\centering
\small
\caption{Statistics of the training and held-out evaluation sets.
Binary evaluation treats \textit{Good} as the positive class and
\textit{Fair}$\cup$\textit{Bad} as the negative class.
Attribution positives are counted at the page level over non-Good pages
and are not mutually exclusive across dimensions.}
\label{tab:data_statistics}
\begin{tabular}{lrr}
\toprule
Split / Category & Samples & \% \\
\midrule
Training set & 169,434 & -- \\
Held-out test set & 17,000 & 100.0 \\
\midrule
\multicolumn{3}{l}{\textit{Page-level labels in the test set}} \\
Good & 6,342 & 37.3 \\
Fair & 2,812 & 16.5 \\
Bad  & 7,846 & 46.2 \\
\midrule
\multicolumn{3}{l}{\textit{Binary grouping used in Sec.~\ref{sec:exp_settings}}} \\
Positive (Good) & 6,342 & 37.3 \\
Negative (Fair $\cup$ Bad) & 10,658 & 62.7 \\
\midrule
\multicolumn{3}{l}{\textit{Attribution positives}} \\
Relevance & 5,636 & 52.9 \\
Quality & 5,742 & 53.9 \\
Diversity & 5,332 & 50.0 \\
Authority & 5,967 & 56.0 \\
Heterogeneous & 5,784 & 54.3 \\
\bottomrule
\end{tabular}
\end{table}

\end{document}